\documentstyle[12pt]{article}
\def\beq{\begin{equation}}
\def\eeq{\end{equation}}
\def\bea{\begin{eqnarray}}
\def\eea{\end{eqnarray}}
\def\bq{\begin{quote}}
\def\eq{\end{quote}}

\def\gappeq{\mathrel{\rlap {\raise.5ex\hbox{$>$}} {\lower.5ex\hbox{$\sim$}}}}

\def\lappeq{\mathrel{\rlap{\raise.5ex\hbox{$<$}} {\lower.5ex\hbox{$\sim$}}}}

\begin{document}
\topmargin -0.5cm
\oddsidemargin -0.3cm
\pagestyle{empty}
\begin{flushright} {HIP-1999-33/TH}
\end{flushright}
\begin{center} {\bf Canonical structure and boundary conditions in Yang-Mills
theory}$^{\dag)}$ \\
\vspace*{1.5cm}  {\bf Christofer Cronstr\"{o}m}$^{*)}$\\
\vspace{0.3cm} Physics Department, Theoretical Physics Division \\ FIN-00014 University of
Helsinki, Finland \\
\vspace{0.5cm}
 
\vspace*{2cm}   {\bf ABSTRACT} \\ \end{center}
\vspace*{5mm}
\noindent 
The canonical structure of pure Yang-Mills theory is analysed in the case when
Gauss' law is satisfied identically by  construction. It is shown that the theory has a
canonical structure in this case, provided one uses a special gauge  condition, which is a
natural generalisation of the Coulomb gauge condition of electrodynamics. The emergence of a
canonical structure depends critically also on the boundary conditions used for the relevant
field variables.  Possible boundary conditions are analysed in detail.  A comparison of the
present formulation in the generalised Coulomb gauge with the well known Weyl gauge ($A_{0} =
0$) formulation  is made. It appears that the  Hamiltonians in these two formulations differ
from one another in a non-trivial way. It is still an open question whether these  differences
give rise to truly different structures upon quantisation. An extension of the formalism to
include coupling to fermionic fields is briefly discussed.

\vspace*{1cm} 
\noindent
 

\begin{flushleft}
$^{\dag)}$ Submitted to the EPS-HEP99 conference in Tampere, Finland, July 1999 
\vspace*{1cm}
\end{flushleft}
$^{*)}$ E-mail: Christofer.Cronstrom@Helsinki.fi

\vfill\eject


\setcounter{page}{1}
\pagestyle{plain}

\section{Introduction}

This paper is concerned with analysing the canonical structure of pure Yang-Mills theory
\cite{YM} along the lines of  two previous publications \cite{Chris 1}, \cite{Chris 2}, in
which a new gauge condition, called the generalized Coulomb  gauge condition  was introduced
and used to obtain a straightforward canonical formulation of Yang-Mills theory, in the  case
when Gauss law is satisfied identically by construction. In the previous papers certain
assumptions were made  concerning the spatial boundary conditions of the Yang-Mills
potentials. The boundary conditions in question are very  important for the elucidation of the
canonical structure. Here I will analyse the requisite boundary conditions in  detail, and
show that there is a set of boundary conditions at spatial infinity, which is consistent with
the  canonical structure. Rigorous proofs and heavy mathematical machinery are omitted here in
the interest of simplicity. 

The basic quantity of Yang-Mills theory is the {\em action} $S$, which is given in terms of
the gauge field 
$G_{\mu \nu}$ as follows,
\begin{equation} S = - \; \frac{1}{4}\; \int d^{4}x (G_{\mu\nu}(A), G^{\mu\nu}(A)).
\label{eq:action}
\end{equation} where the inner product $(\;, \;)$ is defined below. For the gauge field
$G_{\mu \nu}$ I use a matrix notation  (summation over repeated indices),
\begin{equation} G_{\mu\nu}(A) \equiv   G_{\mu\nu}^{~~a}(A)T_{a} = \partial_{\nu}A_{\mu}(x) -
\partial_{\mu}A_{\nu}(x) - ig[A_{\mu}(x), A_{\nu}(x)],
\label{eq:Gfield}
\end{equation} where the quantities $T_{a}$ are matrices in a convenient representation of the
Lie algebra of the gauge group $G$, 
\begin{equation} [T_{a}, T_{b}] = if_{ab}^{~~c}T_{c},
\label{eq:liealg}
\end{equation} and the quantities $A_{\mu}$ are the gauge potential components,
\begin{equation} A_{\mu}(x) = A_{\mu}^{~a}T_{a}.
\label{eq:Apot}
\end{equation}

It is assumed that the gauge group  $G$ is semisimple and compact. The inner product $(\;,
\;)$ for any two Lie algebra valued (matrix valued) quantities $A = A^{a}T_{a}$ and $B =
B^{a}T_{a}$ is then  expressed with the aid of  the Lie algebra structure constants
$f_{ab}^{~~c}$ as follows, 
\begin{equation} (A, B) = h_{ab}A^{a}B^{b},
\label{eq:inprod}
\end{equation} where
\begin{equation} h_{ab} = -f_{ab'}^{~~c'}f_{bc'}^{~~b'}.
\label{eq	:Killing}
\end{equation} The quantity $h_{ab}$ and its inverse $h^{ab}$ are used to lower and rise Lie
algebra indices, respectively. The notation  used here is otherwise pretty standard or
self-explanatory, with e.g. Greek letters used as indices denoting Minkowski  space indices
ranging from $0$ to $3$, and latin indices from the middle of the alphabet $(k, \ell, ...)$
denoting space indices ranging from $1$ to
$3$. The Minkowski space metric is taken to be diagonal, with signature $(+,-,-,-)$. Unless
otherwise stated, repeated indices are always summed over, be they Lie algebra-, spacetime- or
space indices. 

It is convenient for future reference to write the action (\ref {eq:action}) in terms of a
Lagrangian $L$.  The action $S$ is the integral of the Lagrangian $L$ in an appropriate time
interval $[x^{0}_{i}, x^{0}_{f}]$. Thus,
\begin{equation} S = \int_{x^{0}_{i}}^{x^{0}_{f}} dx^{0} L,
\label{eq:defSL}
\end{equation} where 
\begin{equation} L = -\frac{1}{2} \int_{V} d^{3} {\bf x}\left (G_{0k}(A), G^{0k}(A) \right ) -
\frac{1}{4} \int_{V} d^{3} {\bf x}\left (G_{k\ell}(A), G^{k\ell}(A) \right ).
\label{eq:defL}
\end{equation} In the expression (\ref {eq:defL}) for the Lagrangian $L$, the quantity $V$ is
some appropriate domain in ${\bf R}^{3}$, which yet has to be specified.  

As is well known, requiring the action (\ref{eq:action}) to be stationary with respect to
local  variations of {\em all} the potential components $A_{\mu}$, considered as independent
quantities, yields the follwing field equations,
\begin{equation}
\nabla_{\nu}(A) G^{\mu\nu}(A) = 0\;, \mu =  0,1,2,3.
\label{eq:fieldeq}
\end{equation} The "covariant gradient" $\nabla_{\mu}(A)$ used above in Eq. (\ref {eq:fieldeq}) 
is a convenient  notion,      
\begin{equation}
\nabla_{\mu}(A) \equiv \partial_{\mu} + ig\,[A_{\mu}, \;\;\;],
\label{eq:covgrad}
\end{equation} which will be frequently used in what follows. 
 
The non-Abelian Gauss law is obtained from the equations (\ref {eq:fieldeq}) for $\mu = 0$. 
Expressed in terms of the potential $A$ the non-Abelian Gauss law reads as follows, 
\begin{equation}
\nabla_{k}(A)\nabla^{k}(A) A^{0} -  \nabla_{k}(A){\dot A}^{k} = 0,
\label{eq:nAGauss}
\end{equation}
where 
\begin{equation}
\dot{A}_{k}(x) \equiv \partial_{0}A_{k}(x).
\label{eq:dot}
\end{equation}

The time derivative of any quantity will frequently in what follows be denoted by a dot on top
of that quantity, as in the equation (\ref {eq:dot}) above.

\subsection{The Hamiltonian or Weyl gauge formalism}

It is well known \cite{Jackiw} that Yang-Mills theory can be expressed in a canonical form in
the  Hamiltonian gauge, or so-called Weyl gauge \cite{Weyl} $A_{0} = 0$.  

The Yang-Mills Lagrangian in the case $A_{0} = 0$, which will be called $L_{W}$ here, is
obtained from the expression (\ref {eq:defL}) by putting $A_{0} = 0$ in that expression, 
\begin{equation} L_{W} = - \frac{1}{2} \int d^{3}{\bf x} (\dot{A}_{k}, \dot{A}^{k}) -
\frac{1}{4} \int d^{3}{\bf x}(G_{k\ell}, G^{k\ell}).
\label{eq:LWeyl}
\end{equation} The Lagrangian (\ref {eq:LWeyl}) describes a theory which is not quite the same
as Yang-Mills theory, since Gauss' law is absent from  the field equations following from the
action principle with the expression  (\ref {eq:LWeyl}) as Lagrangian. Gauss' law takes the
following form in the case when $A_{0} = 0$, as seen  from Eq. (\ref {eq:nAGauss}) above,
\begin{equation}
\nabla_{k}(A) \dot{A}^{k} = 0.
\label{eq:GWeyl}
\end{equation} The usual way to analyse Yang-Mills theory in the Weyl gauge $A_{0} = 0$, is to
proceed from the Lagrangian (\ref {eq:LWeyl}), disregarding Gauss' law to begin with. Using
the variables $A_{k}^{a}$ and $\dot{A}_{k}^{a}$ as  generalised coordinates and velocities, 
respectively, it is then perfectly simple to derive a canonical Hamiltonian  formalism for the
system defined by the Lagrangian (\ref {eq:LWeyl}). The corresponding Hamiltonian $H_{W}$ is,
\begin{equation} H_{W} =  - \frac{1}{2} \int d^{3}{\bf x} (\pi_{k}, \pi^{k}) + \frac{1}{4}
\int d^{3}{\bf x}(G_{k\ell}, G^{k\ell}),
\label{eq:HamWeyl}    
\end{equation} 
i.e. a simple sum of a kinetic term, depending on canonical momenta
$\pi^{k}_{a}$ only, and a potential, or interaction term,  depending on conjugate canonical
coordinates $A_{k}^{a}$ only. 

\section{First attempt at a canonical formulation when Gauss' law is in force}

Gauss' law, Eq. (\ref {eq:nAGauss}) above, will now be considered as an equation determining
the  (matrix valued) potential component $A_{0}$, for given space components ${\bf A}$ and 
${\bf \dot {A}}$. This equation is a system of {\em linear, elliptic partial differential
equations} with time $x^{0}$ acting as a parameter in an appropriate interval, with
${\bf x} \in V$ being the independet variables. I will discuss the solvability of Eq.  (\ref
{eq:nAGauss}) subsequently, but proceed now by assuming the existence of unique solution 
$A_{0}$, which is a {\em functional}  of the space components ${\bf A}$ and their time
derivatives
$\partial_{0}{\bf A} \equiv {\dot{{\bf A}}}$, i.e.
\begin{equation} A_{0} = A_{0}\left \{{\bf A}, {\dot{{\bf A}}}\right \}.
\label{eq:Aofunct}
\end{equation}

The question is then whether the Yang-Mills system, which is originally defined by the action 
(\ref {eq:action}), permits a canonical structure when the potential component $A_{0}$ is a 
solution to Gauss' law (\ref {eq:nAGauss}), i.e. when $A_{0}$ is given in terms of ${\bf A}$
and 
${\dot{{\bf A}}}$ by the expression (\ref {eq:Aofunct}). It is possible to get some insight
into this question without specifying the actual functional form of the relation (\ref
{eq:Aofunct}) in minute  detail.

The Lagrangian of the Yang-Mills system, when the potential component $A_{0}$, is given by the 
relation (\ref {eq:Aofunct}) above, is  obtained simply by inserting the solution (\ref
{eq:Aofunct}) for $A_{0}$ into the Lagrangian (\ref {eq:defL}) above. The resulting Lagrangian 
will be called $L_{0}$, and is explicitly given as follows,
\begin{eqnarray}
\label{eq:L_{0}}
 L_{0} & = &   - \frac{1}{2}\int_{V} d^{3} {\bf x} \left (\nabla_{k}({\bf A})
A_{0}\left \{{\bf A}, {\dot{{\bf A}}}\right \}
 - \dot{{\bf A}}_{k}, \nabla^{k}({\bf A}) A^{0}\left \{{\bf A}, {\dot{{\bf A}}}\right \} -
\dot{{\bf A}}^{k}
\right ) \\ 
 & & - \frac{1}{4} \int_{V} d^{3} {\bf x} \left ( G_{kl}({\bf A}), G^{kl}({\bf A})
\right ).  \nonumber
\end{eqnarray} 

At this point it is appropriate to check whether the action principle involving
the Lagrangian $L_{0}$ in (\ref {eq:L_{0}}) above reproduces the field equations (\ref
{eq:fieldeq}). It is perfectly straightforward to verify the following result,
\begin{eqnarray}
\label{eq:varL_{0}}
\delta\;\int_{x^{0}_{i}}^{x^{0}_{f}} dx^{0} L_{0} & = & - \int_{x^{0}_{i}}^{x^{0}_{f}} dx^{0}
\int_{V}d^{3}{\bf x} \left (\delta A_{k}, \nabla_{0}(A)(\nabla^{k}(A) A^{0}\left \{{\bf A},
{\dot{{\bf A}}}\right \} - \dot{{\bf A}}^{k}) - \nabla_{\ell}G^{k\ell}(A) \right ) \\ &
& -
\int_{x^{0}_{i}}^{x^{0}_{f}} dx^{0} \int_{\partial V} d^{2}\sigma_{k} \left (\delta A^{0}\left \{{\bf A}, {\dot{{\bf
A}}}\right \}, \nabla^{k}(A)A^{0}\left \{{\bf A}, {\dot{{\bf A}}}\right \} - \dot{{\bf
A}}^{k} \right ). \nonumber
\end{eqnarray}

Now the boundary conditions for the solution $A_{0}$ to Gauss' law, i.e. the system of linear
elliptic partial differential equations (\ref {eq:nAGauss}), enter into the discussion. If the
surface term  in Eq. (\ref {eq:varL_{0}}) is  non-vanishing, and not by itself a variation of
some surface functional, then the Lagrangian (\ref {eq:L_{0}}) is not a valid Lagrangian in
the action principle which  is supposed to lead to the field equations (\ref {eq:fieldeq}) for
$\mu = 1,2,3$, when $A_{0}$ is  given by (\ref {eq:Aofunct}).

In the first place the domain $V$ is actually considered to be all of ${\bf R}^{3}$. The
integrals  over $V$ will be given the following interpretation,
\begin{equation}
\int_{V} d^{3}{\bf x} \cdot \cdot \cdot = \lim_{R \rightarrow \infty} \int_{\mid {\bf x} \mid
< R} d^{3}{\bf x} \cdot \cdot \cdot 
\label{eq:intV}
\end{equation} The vanishing of the surface term in Eq. (\ref {eq:varL_{0}}) is then
equivalent to the follwing,
\begin{equation}
\lim_{R \rightarrow \infty} \int_{\mid{\bf x} \mid = R} d\Omega R^{2}\left (\delta A^{0}[{\bf
A}, \dot{{\bf A}}], \nabla^{(r)}(A)A^{0}\left \{{\bf A}, {\dot{{\bf A}}}\right \} - \dot{{\bf
A}}^{(r)} \right ) = 0,
\label{eq:surt}
\end{equation} 
where the superscript $(r)$ denotes the {\em radial} component of the
corresponding quantity.

Thus, {\em if } the surface term (\ref {eq:surt}) vanishes for all admissible variations of
the  independent generalised coordinates ${\bf A}$ and velocities $\dot{{\bf A}}$,
respectively,  then   the variational principle
\begin{equation}
\delta\;\int_{x^{0}_{i}}^{x^{0}_{f}} dx^{0} L_{0} = 0,
\label{eq:var2}
\end{equation} leads to the following equations of motion,
\begin{equation}
\nabla_{0}(A)(\nabla^{k}(A) A^{0}\left \{{\bf A}, {\dot{{\bf A}}}\right \} - \dot{{\bf
A}}^{k}) - \nabla_{\ell}G^{k\ell}(A) = 0,
\label{eq:eqmo2}
\end{equation} as is evident from the relation (\ref {eq:varL_{0}}). Needless to say, the
equations (\ref {eq:eqmo2}) are nothing but the field  equations (\ref {eq:fieldeq}) for $\mu
= 1,2,3$, with $A_{0}$ given by the  formal solution (\ref {eq:Aofunct}) to Gauss' law (\ref
{eq:nAGauss}).  The vanishing of the  surface term (\ref {eq:surt}) depends on the assumed
asymptotic behaviour of the {\em independent} variables ${\bf A}$ and ${\bf \dot{A}}$, as well
as on the boundary conditions (at spatial infinity) of the {\em dependent} variable $A_{0}$. I
will return to this question below, and continue for the time being by assuming that the
relation (\ref {eq:surt}) is valid. Then the Lagrangian $L_{0}$ given in (\ref {eq:L_{0}})
ought to be a suitable starting point for the construction of a Hamiltonian and the
corresponding canonical variables by means of a Legendre transform in the usual way.

The formal definition of the canonical momentum $P_{k}^{a}$ conjugate to the coordinate 
$A_{a}^{k}$ is,
\begin{equation} P_{k}^{a}(x^{0}, {\bf x}) \equiv \frac{\delta L_{0}}{\delta
\dot{A}_{a}^{k}(x^{0}, {\bf x})} = \left ( \nabla_{k}(A)A_{0}\left \{{\bf A}, {\dot{{\bf
A}}}\right \} \right )^{a}(x^{0}, {\bf x}) - \dot {A}_{k}^{a}(x^{0}, {\bf x}),
\label {eq:canmom1}
\end{equation} where the condition (\ref {eq:surt}) has been used in the calculation of the
functional derivative of $L_{0}$ in (\ref {eq:canmom1}) above. In view of the fact that
$A_{0}$ in Eq. (\ref {eq:canmom1}) satisfies Gauss' law (\ref {eq:nAGauss}), one finds
immediately from (\ref {eq:canmom1}) that
\begin{equation}
\nabla^{k}(A)P_{k}(x^{0}, {\bf x}) \equiv 0.
\label{eq:divP=0}
\end{equation} Now one is supposed to be able to solve Eq. (\ref {eq:canmom1}) for the
generalised velocity
$\dot{A}_{k}^{a}$ in terms of ${\bf A}$ and ${\bf P}$, respectively. But this is impossible, 
since, Eq. (\ref {eq:canmom1}) can not be solved for the quantity $\Gamma$ defined below,
\begin{equation}
\Gamma(x^{0}, {\bf x}) \equiv \nabla_{k}(A)\dot{A}^{k}(x^{0}, {\bf x}),
\label{eq:defGamma}
\end{equation} i.e. if one tries to derive an equation for the quantity $\Gamma$ defined
above from Eq. (\ref {eq:canmom1}), one gets a completely vacuous identity for this quantity,
as a result of Eq. (\ref {eq:divP=0}).

It would seem then, that the canonical formalism breaks down for the case at hand. However
this is  not necessarily the case. The difficulty described above can be avoided if one can
manage to make
$A_{0}$ {\em independent} of the generalized velocity variables ${\bf \dot{A}}$. This can be 
accomplished in the present situation ($A_{0} \neq 0$) by imposing the following  {\em gauge
condition},
\begin{equation}
\nabla_{k}(A)\dot{A}^{k}(x^{0}, {\bf x}) = 0.
\label{eq:genCg}
\end{equation} 
The condition (\ref {eq:genCg}) is the {\em generalized Coulomb gauge
condition} referred to  previously. The fact that this is actually a {\em proper gauge
condition}, has been demonstrated in Ref. \cite {Chris 1}.

However, if one uses the gauge condition (\ref {eq:genCg}) then  the generalised velocities 
$\dot{A}^{k}$  are no longer  independent quantities, and then one cannot use the formula 
(\ref {eq:canmom1}) as it stands for the construction of the canonical momentum variables. An 
alternative procedure which leads to a proper canonical formalism will be given below.

\section{Gauss' law and asymptotic conditions}

I now assume that the generalized Coulomb gauge condition (\ref {eq:genCg} is in force. Then Gauss' law,
Eq. (\ref {eq:nAGauss}), takes the following form,
\begin{equation}
\nabla_{k}(A)\nabla^{k}(A) A^{0} = 0.
\label{eq:n0AGauss}
\end{equation}
If one demands that the solution to Eq. (\ref {eq:n0AGauss}) vanishes roughlty speaking faster than 
$\mid {\bf x} \mid ^{-\frac{1}{2}}$ for $\mid {\bf x}\mid \rightarrow \infty$, then the solution vanishes identically.
This can be seen as follows. Take the inner product of Eq. (\ref {eq:n0AGauss}) with $A_{0}$ and integrate over 
$ {\bf x}$. Using the ordinary divergence theorem one readily obtains the following result,
\begin{equation}
\lim_{R \rightarrow \infty}\; \int_{\mid {\bf x}\mid < R} d^{3}{\bf x} (\nabla_{k}(A) A_{0}, \nabla^{k}(A) A_{0})
= \lim_{R \rightarrow \infty}\; \int_{\mid {\bf x}\mid = R} d\Omega R^{2} (A_{0},
\nabla^{(r)}(A)A_{0}).
\label{eq:surf2}
\end{equation}
Assuming now
\begin{equation}
A_{0}(x^{0}, {\bf x})\mid_{\mid {\bf x} \mid = R}\; \sim R^{-\gamma}\;,\;\; \nabla^{(r)}(A) A_{0}(x^{0}, {\bf
x})\mid_{\mid {\bf x} \mid = R}\; \sim R^{-\gamma - 1}\;, \gamma > \frac{1}{2},
\label{eq:asyAo}
\end{equation}
one finds that the limiting value of the right hand side of Eq. (\ref {eq:surf2}) is zero. Since the inner product
$(\;, \;)$ is positive definite, one then concludes from Eq. (\ref {eq:surf2}) that $A_{0}$ is {\em covariantly
constant}. However, since $A_{0}$ vanishes at infinity by assumption, the covariant constant is zero,
i.e. that
\begin{equation}
A_{0}(x^{0}, {\bf x}) \equiv 0.
\label{eq:Azerop}
\end{equation}
Needless to say, the asymptotic condition (\ref {eq:asyAo}) above can be somewhat refined; all that is needed to 
obtain the result (\ref {eq:Azerop}) is that the right hand side of Eq. (\ref {eq:surf2}) vanishes, which guarantees
that $A_{0}$ is covariantly constant, and then that $A_{0}$ has the limiting value $0$ at infinity (or at
some finite point), so that the covariant constant in question actually vanishes. 

However, the class of functions with (roughly speaking) the asymptotic behaviour (\ref {eq:asyAo}) does not exhaust 
the class of possible solutions to  Eq. (\ref {eq:n0AGauss}). It is also possible to consider functions $A_{0}$ which
approach a non-vanishing {\em constant} matrix at space infinity,
\begin{equation}
\lim_{\mid {\bf x} \mid \rightarrow \infty} A_{0}(x^{0}, {\bf x})  = \Lambda \equiv \Lambda^{a}T_{a},
\label{eq:Aolambda}
\end{equation}
where the real quantities $\Lambda^{a}$ are absolute constants. In addition to Eq. (\ref {eq:Aolambda}) one can
impose an asymptotic condition on the radial derivative of $A_{0}$,
\begin{equation}
\lim_{\mid {\bf x} \mid \rightarrow \infty} \mid {\bf x} \mid \frac{\partial}{\partial \mid {\bf x} \mid}
A_{0}(x^{0}, {\bf x}) = 0.
\label{eq:asym3}
\end{equation}
Requiring the validity of the asymptotic conditions (\ref {eq:Aolambda}) and (\ref {eq:asym3}) and using standard
arguments of potential theory \cite{Kellogg}, one now derives an integral representation  for the function $A_{0}$
involving the (ordinary) Laplacian of that function,
\begin{equation}
A_{0}(x^{0}, {\bf x}) = \Lambda - \frac{1}{4\pi} \int_{{\bf R}^{3}} d^{3}{\bf y} \frac{1}{\mid{\bf x} - {\bf y} \mid}
\nabla_{\bf y}^{2} A_{0}(x^{0}, {\bf y}),
\label{eq:irepr1}
\end{equation}
Using the the present form of Gauss' law, Eq. (\ref {eq:n0AGauss}), one then converts (\ref {eq:irepr1}) into an
integral equation for the determination of $A_{0}$. For this purpose it is convenient to introduce some new notation,
\begin{equation}
U^{ka}_{~~~b}(x^{0}, {\bf y}) := 2gf^{a}_{~~bc} A^{kc}(x^{0}, {\bf y}),
\label{eq:def U}
\end{equation}
and
\begin{equation}
V^{a}_{~~b}(x^{0}, {\bf y}) := 2gf^{a}_{~~bc} \frac{\partial}{\partial y^{k}} A^{kc}(x^{0}, {\bf y}) +
g^{2}f^{a}_{~~c'd}f^{c'}_{~~ba'}A_{k}^{~~d}(x^{0}, {\bf y})A^{ka'}(x^{0}, {\bf y}).
\label{eq:def V}
\end{equation}
One then obtains the following integral equations,
\begin{equation}
A_{0}^{a}(x^{0}, {\bf x}) = \Lambda^{a} - \frac{1}{4\pi} \int_{{\bf R}^{3}} d^{3}{\bf y} \frac{1}{\mid{\bf x} -{\bf y}
\mid} \left \{U^{ka}_{~~~b}(x^{0}, {\bf y}) \frac{ \partial A_{0}^{b}(x^{0}, {\bf y})}{\partial y^{k}} +
V^{a}_{~~b}(x^{0}, {\bf y}) A_{0}^{b}(x^{0}, {\bf y}) \right \}.
\label{eq:irepfin}
\end{equation}
The integral equations (\ref {eq:irepfin}) constitute the starting point for the proof of existence of solutions
to the present form of Gauss' law, Eq. (\ref {eq:n0AGauss}). For this one needs naturally also to specify
conditions on the potential components $A_{k}^{a}$, which, together with their space derivatives
determine the (unique) solution to Eq. (\ref {eq:irepfin}). All this is a part of the "heavy
mechinery" mentioned in the Introduction, which  I will omit in this paper. However it is appropriate
to note the following asymptotic conditions, which are needed for the existence of  solutions
$A_{0}^{a}$ to Eq. (\ref {eq:irepfin}),
\begin{equation}
A_{k}^{a}(x^{0}, {\bf x}) \; \sim \frac{1}{\mid {\bf x}\mid^{1 + \epsilon}}\;, \;
\frac{\partial}{\partial x^{\ell}}A_{k}^{a}(x^{0}, {\bf x}) \; \sim \frac{1}{\mid {\bf x}\mid^{2 + \epsilon}}\;, \;
\epsilon > 0.
\label{eq:asym4}
\end{equation}

I will now denote the solution of the system of integral equations (\ref {eq:irepfin}) by 
$A_{0}\left \{{\bf A}\right \}$, and  the corresponding Lagrangian by $L_{00}$ (compare with
Eq. (\ref {eq:L_{0}})) ,
\begin{eqnarray}
\label{eq:L_{00}}
 L_{00} & = &   - \frac{1}{2}\int_{V} d^{3} {\bf x} \left (\nabla_{k}({\bf A}) A_{0}\left \{{\bf A}\right \}
 - \dot{{\bf A}}_{k}, \nabla^{k}({\bf A}) A^{0}\left \{{\bf A}\right \} - \dot{{\bf A}}^{k} \right ) \\ 
 & & - \frac{1}{4} \int_{V} d^{3} {\bf x} \left ( G_{kl}({\bf A}), G^{kl}({\bf A}) \right ). \nonumber
\end{eqnarray}
The Lagrangian (\ref {eq:L_{00}}) will now in the next section be used to derive the Hamiltonian for
the Yang-Mills system under the condition that the generalized Coulomb gauge (\ref {eq:genCg}) is in
force. I implement this condition as a {\em constraint}, by means of a (matrix valued) Lagrange
multiplier field $C(x)$, which is used to modify the Lagrangian (\ref {eq:L_{00}}) as follows,
\begin{equation} L_{00} \rightarrow L' = L_{00} + \int_{V} d^{3}{\bf x} (C(x),
\nabla_{k}(A)\dot{A}^{k}).
\label{eq:Lconstr}
\end{equation} 

\section{Canonical coordinates, momenta and Hamiltonian}

I now make a direct transition to a Hamiltonian formulation using the modified Lagrangian  (\ref
{eq:Lconstr}) above, in a manner described in the general case by Berezin \cite{Berezin}.  The
starting point is the familiar definition of canonical momentum variables $\pi_{k}^{a}$,
\begin{equation}
\pi_{k}^{a}(x^{0}, {\bf x}) \equiv \frac{\delta L'}{\delta \dot{A}_{a}^{k}(x^{0}, {\bf x})} =
\left (\nabla_{k}(A) A_{0}\left \{{\bf A} \right \} \right )^{a} - \dot{A}_{k}^{a} - \left (
\nabla_{k}(A) C \right )^{a},
\label{eq:canpi}
\end{equation}
The equations (\ref {eq:canpi}) above, {\em together} with the constraint equations (\ref {eq:genCg}) are now
supposed to be solved for the quantities $\dot{A}_{k}^{a}$ and $C^{a}$ in terms of the canonical coordinates
$A_{a}^{k}$ and momenta  $\pi_{k}^{a}$, respectively. Using Eqns. (\ref {eq:genCg}) and (\ref {eq:n0AGauss}),
one finds immediately from Eq. (\ref {eq:canpi}) that
\begin{equation}
 - \nabla_{k}(A)\nabla^{k}(A) C = \nabla^{k}(A)\pi_{k},
\label{eq:defCbc}
\end{equation} 
which, together with appropriate boundary conditions, defines the quantity $C$ as an $x$-dependent functional 
of ${\bf A}$ and ${\bf \pi}$, respectively,
\begin{equation} 
C  =  C\left \{ {\bf A}, {\bf \pi} \right \}(x^{0}, {\bf x}).
\label{eq:defC2}
\end{equation} 
It is certainly desirable that the solution $C\left \{ {\bf A}, {\bf \pi} \right \}$ to
Eq. (\ref {eq:defCbc}) be {\em unique}. The uniqueness is guaranteed if one uses the following
boundary condition,
\begin{equation}
\lim_{R \rightarrow \infty}\; \int_{\mid {\bf x}\mid = R} d\Omega R^{2} (C(x), \nabla^{(r)}(A)C(x)) =
0.
\label{eq:boundcC}
\end{equation}
The proof of uniqueness of the solution to Eq. (\ref {eq:defCbc}) under the condition (\ref {eq:boundcC})
has essentially already been given above in connection with Eq. (\ref {eq:n0AGauss}). Namely, if
there are two distinct solutions to Eq. (\ref {eq:defCbc}), then their difference satisfies the
corresponding homogeneous equation, which is precisely of the form (\ref {eq:n0AGauss}). However,
under the condition  (\ref {eq:boundcC}), the homogeneous equation in question has only the trivial
zero solution, as demonstrated  in the discussion following Eq.  (\ref {eq:n0AGauss}). Hence the
solution  $C\left \{ {\bf A}, {\bf \pi} \right \}$ is unique. 

One now straightforwardly expresses the generalised velocity in terms of coordinate- and
momentum  variables,
\begin{equation}
\dot{A}_{k}^{a} = \left (\nabla_{k}(A_{0}\left \{{\bf A} \right \} - C) \right )^{a} -
\pi_{k}^{a}.
\label{eq:defAdot}
\end{equation} 
The construction of the Hamiltonian $H$ then proceeds in the usual way. The
relation defining the Hamiltonian $H$ is the following,
\begin{equation} H = \int_{V} d^{3}{\bf x} (\pi_{k}, \dot{A}^{k}) - L_{00},
\label{eq:defH}
\end{equation} where the quantity $\dot{A}^{k}$ ocurring in the expressions in (\ref
{eq:defH}) should be  given in terms of canonical variables by the expression (\ref
{eq:defAdot}). It should be observed, that it is indeed the Lagrangian $L_{00}$ which enters
in the definition of the  Hamiltonian $H$ above, since at this stage the constraint (\ref
{eq:genCg}) is an identity.

By straightforward calculation one finally obtains the Hamiltonian expressed in terms of
canonical  variables from the definition (\ref {eq:defH}),  
\begin{eqnarray}
\label{eq:finHam} 
H & = & -\frac{1}{2} \int_{V}d^{3}{\bf x} \left (\pi_{k}, \pi^{k} \right ) +
\frac{1}{4} \int_{V} d^{3} {\bf x} \left ( G_{kl}(A), G^{kl}(A) \right )  \\ & & +
\int_{V}d^{3}{\bf x} \left (\pi_{k}, \nabla^{k}(A)A^{0}\left \{{\bf A} \right \} \right ) +
\frac{1}{2} \int_{V}d^{3}{\bf x} \left (\nabla_{k}(A)C, \nabla^{k}(A)C \right ), \nonumber
\end{eqnarray} 
where the quantity $C$ is the appropriate  solution to the system of linear
elliptic partial  differential equations (\ref {eq:defCbc}), as discussed previously.

\section{The Hamiltonian equations of motion}

The Hamiltonian equations of motion are obtained by functional differentiation of the Hamiltonian
(\ref {eq:finHam}) with respect to the canonical momenta and coordinates, respectively.  

The  equations of motion for the coordinates are as follows,
\begin{equation}
\dot{A}_{k}^{a}(x^{0}, {\bf x}) \equiv \frac{\delta H}{\delta \pi_{a}^{k}(x^{0}, {\bf x})} =  
\left (\nabla_{k}(A)(A_{0}\left \{{\bf A}\right \} - C)\right )^{a} - \pi_{k}^{a},
\label{eq:dHdpi}
\end{equation} 
which agree  precisely with the expressions (\ref {eq:defAdot}) as they should. In the calculation
leading to Eq. (\ref {eq:dHdpi}) one encounters the following surface term,
\begin{equation}
\lim_{R \rightarrow \infty} \int_{\mid{\bf x} \mid < R} d^{3}{\bf x} \partial_{k} (C,
\nabla^{k}(A)\delta_{\pi}C) = \lim_{R \rightarrow \infty} \int d\Omega R^{2} (C, \nabla^{(r)}(A)
\delta_{\pi}C),
\label{eq:surfpi}
\end{equation}
which must vanish in order that the functional derivative in Eq. (\ref {eq:dHdpi}) be well defined. The surface
term (\ref {eq:surfpi}) in question vanishes as it should, in view of the general boundary conditions (\ref
{eq:boundcC}) imposed on the quantity $C\left \{ {\bf A}, {\bf \pi} \right \}$. 

The calculation of the functional derivative of the Hamiltonian (\ref {eq:finHam}) with respect to the
generalized coordinate variables $A_{k}^{a}$ is fairly lengthy, but nevertheless straightforward.
In the course of the  calculation one finds that a surface term analoguous to Eq. (\ref {eq:surt}) 
has to vanish in  order that the functional derivative in question be well defined, i.e. that  
\begin{equation}
\lim_{R \rightarrow \infty} \int_{\mid{\bf x} \mid = R} d\Omega R^{2}\left (\delta A^{0}[{\bf
A}], \nabla^{(r)}(A)A^{0}\left \{{\bf A} \right \} - \dot{{\bf A}}^{(r)} \right ) = 0.
\label{eq:surfA_{k}}
\end{equation}
The condition (\ref {eq:surfA_{k}}) is quite non-trivial. It should be remembered, that the variation of the
quantity $A_{0}$ can not be declared to vanish outside some finite region, since $A_{0}$ is a dependent
variable, determined by Gauss' law,  i.e. in the present case by Eq. (\ref {eq:irepfin}). Under
essentially the conditions (\ref {eq:asym4}) with $\epsilon > 0$, I have been able to prove, that the
iterative solution of the Eqns. (\ref {eq:irepfin}), which are equivalent to the system of  partial
differential equations (\ref {eq:n0AGauss}) with the boundary conditions (\ref {eq:Aolambda}) and
(\ref {eq:asym3}), is such that, for large $\mid {\bf x} \mid$, 
\begin{equation}
\delta A_{0}({\bf A})(x^{0}, {\bf x}) = O \left (\frac{1}{\mid {\bf x} \mid } \right ),
\label{eq:deltAasy}
\end{equation}
for any {\em local} variations $\delta {\bf A}$. If one assumes that $\epsilon > 1$ in the discussion above,
it is essentially trivial to show conclusively that Eq. (\ref {eq:deltAasy}) is valid. However, I will proceed
by assuming that Eq. (\ref {eq:deltAasy}) is valid also if one merely takes $\epsilon > 0$, which is a
plausible assumption as discussed  above. 

If the conditions (\ref {eq:asym4}) and  (\ref {eq:deltAasy}) are valid, or more precisely if the
condition (\ref {eq:surfA_{k}}) is in force, then one obtains,
\begin{eqnarray}
\label{eq:dHdA}
\dot{\pi}_{a}^{k}(x^{0}, {\bf x}) \equiv - \frac{\delta H}{\delta A_{k}^{a}(x^{0}, {\bf x})}&
= & -ig[A_{0}\left \{{\bf A}  \right \} - C, \nabla^{k}(A)C + \pi^{k}]_{a}  \\ & & + \left
(\nabla_{\ell}G^{k\ell}(A) \right )_{a} - ig[C, \nabla^{k}(A)A_{0}\left \{{\bf A}  \right
\}]_{a}. \nonumber
\end{eqnarray}

The pairs of equations, (\ref {eq:dHdpi}) and (\ref {eq:dHdA}) are supposed to be equivalent to the
original field equations (\ref {eq:fieldeq}). I will analyse this equivalence below. For this purpose 
it is convenient to note that the equations (\ref {eq:dHdpi}) and (\ref {eq:dHdA}) admit a {\em constant of
motion},
\begin{equation}
\partial_{0}(\nabla_{k}(A) \pi^{k}) = 0,
\label{eq:como}
\end{equation}
which is crucial for the establishing of the equivalence between Eqns. (\ref {eq:fieldeq}) and the
Hamiltonian equations of motion  (\ref {eq:dHdpi}) and (\ref {eq:dHdA}).  Recalling the equation
(\ref {eq:defCbc}) defining the quantity $C$ together with the boundary conditions (\ref {eq:boundcC}), one
finds using Eq. (\ref {eq:como}) that 
\begin{equation}
\nabla_{k}(A)\nabla^{k}(A) C = K_{1},
\label{eq:K1const}
\end{equation}
where $K_{1}$ denotes a constant.

So, if the equations of motion for the canonical variables $A$ and $\pi$, respectively, are in force, then the 
quantity $C$ is determined by the equation (\ref {eq:K1const}), where the constant $K_{1}$ is so far unknown.

Eliminating the canonical momentum variables $\pi$ between the equations (\ref {eq:dHdpi}) and (\ref
{eq:dHdA}) one obtains the following equations,
\begin{equation}  
\nabla_{\nu}(A)G^{k\nu}(A) + \nabla^{k}(A)\dot{C} + 2ig \left [ \dot{A}^{k}, C \right ] = 0\,,\; k =
1,2,3.
\label{eq:L'-eqns}
\end{equation}
At this point one should recall that Gauss' law is in force by construction,
\begin{equation}
\nabla_{k}(A) G^{0k}(A) = 0.
\label{eq:Gauss5}
\end{equation}
Were it not for the terms involving the quantity $C$ in Eqns. (\ref {eq:L'-eqns}) one could now conclude that
the Hamiltonian equations of motion (\ref {eq:dHdpi}) and (\ref {eq:dHdA}) are equivalent to the original
field equations (\ref {eq:fieldeq}). In order to obtain the required equivalence one must demand that
\begin{equation}
\nabla^{k}(A)\dot{C} + 2ig \left [ \dot{A}^{k}, C \right ] = 0, 
\label{eq:Cequmo}
\end{equation}
simultaneously with the Hamiltonian equations of motion. In this situation the quantity $C$ is determined by
the partial differential equations (\ref {eq:K1const}). However, it does not appear to be possible to prove
that the unique solution to (\ref {eq:K1const}) also satisfies (\ref {eq:Cequmo}) if the constant $K_{1}$ has an
arbitrary value, different from zero. However if
\begin{equation}
K_{1} = 0,
\label{eq:K1zero}
\end{equation}
then substituting this value in Eq. (\ref {eq:K1const}) one obtains the final equation determining the quantity
$C$, when all the other equations of motion are in force,
\begin{equation}
\nabla_{k}(A)\nabla^{k}(A) C = 0.
\label{eq:finC0}
\end{equation}
But Eq. (\ref {eq:finC0}) has only the trivial solution $C = 0$ in the class of functions satisfying the
boundary conditions (\ref {eq:boundcC}) as discussed previously. Hence if the constant of integration $K_{1}$ in
Eq. (\ref {eq:K1const})  equals zero, then Eq. (\ref {eq:Cequmo}) is trivially true, whence the
Hamiltonian equations (\ref {eq:dHdpi}) and (\ref {eq:dHdA})  are completely equivalent to the
original field equations (\ref {eq:fieldeq}). Thus by demanding that the condition (\ref {eq:K1zero})
be valid, one obtains {\em complete equivalence} between the original field equations (\ref
{eq:fieldeq}) and the Hamiltonian equations of motion (\ref {eq:dHdpi}) and (\ref {eq:dHdA}) which
have been obtained using the generalized Coulomb gauge condition (\ref {eq:genCg}).

It is now appropriate to return to the Hamiltonian (\ref {eq:finHam}). This Hamiltonian differs in a 
non-trivial way from the Hamiltonian in the Weyl gauge, i.e. from the expression (\ref
{eq:HamWeyl}), mainly due to the $C$ -dependent terms in the former expression, but also due to the
fact that a non-trivial $A_{0}$-dependence is possible in the case of the Hamiltonian (\ref
{eq:finHam}). The fact that the functional $C\left \{{\bf A}, \pi \right \}$ actually becomes zero
when all the equations of motion are in force, does not mean that one can set $C$ equal to zero in 
the Hamiltonian (\ref {eq:finHam}) and use the resulting expression to generate the appropriate
equations of motion by means of functional differentiation. The {\em functional} dependence of
the quantity $C\left \{{\bf A}, \pi \right \}$ in the Hamiltonian (\ref {eq:finHam}) is essential
in order to generate the proper equations of motion.

Finally I comment briefly on the possibilities to generalize the considerations in this paper to a
situation in which one couples the Yang-Mills field to e.g. a fermionic field. The field
equations (\ref {eq:fieldeq}) then get replaced by the following,
\begin{equation}
\nabla_{\nu}(A)G^{\mu \nu}(A) = J^{\mu} \equiv g\bar{\psi} \gamma^{\mu} T_{a} \psi T^{a},
\label{eq:fermf}
\end{equation}
to which one has to add the field equations for the fermionic fields. Also in this case is it
possible to use the generalized Coulomb gauge condition (\ref {eq:genCg}) so that the
Gauss' law takes the form
\begin{equation}
\nabla_{k}(A) \nabla^{k}(A) A_{0} = J_{0}.
\label{eq:newGauss}
\end{equation}
    
The construction of the canonical variables and Hamiltonian in this case, along similar lines to
those presented here for the pure Yang-Mills case, does not meet with any difficulties of principle.
Likewise, coupling  the Yang-Mills field to a scalar field also gives rise to a system which has a
canonical structure. Details of these constructs as well as some mathematical detail only briefly
touched upon in this paper will be given in future publications \cite {Chris 5}.\\

\vspace*{0.5cm}

\begin{center} 
{\bf Acknowledgements} \\ \end{center}
\noindent
I am indebted to Professor R. Jackiw for useful comments and to Professor B. Koskiaho for  
useful discussions during the course of the work presented here.

\end{document}